\documentclass{llncs}
\usepackage{graphicx}
\usepackage{floatrow}
\usepackage{wrapfig}
\graphicspath{ {./images/} }

\begin{document}

    \pagestyle{empty}
    \mainmatter

    \title{Multi-criteria optimization of workflow-based assembly tasks in manufacturing}

    \author{Florian Holzinger\thanks{corresponding author, \url{florian.holzinger@fh-hagenberg.at}} \and Andreas Beham  
     }

    \institute {
        Heuristic and Evolutionary Algorithms Laboratory\\
        School of Informatics, Communications and Media\\
        University of Applied Sciences Upper Austria\\
        Softwarepark 11, 4232 Hagenberg Austria \\
    }

    \maketitle

\begin{abstract}
Industrial manufacturing is currently amidst it's fourth great revolution, pushing towards the digital transformation of production processes. One key element of this transformation is the formalization and digitization of processes, creating an increased potential to monitor, understand and optimize existing processes. However, one major obstacle in this process is the increased diversification and specialisation, resulting in the dependency on multiple experts, which are rarely amalgamated in small to medium sized companies. To mitigate this issue, this paper presents a novel approach for multi-criteria optimization of workflow-based assembly tasks in manufacturing by combining a workflow modeling framework and the HeuristicLab optimization framework. For this endeavour, a new generic problem definition is implemented in HeuristicLab, enabling the optimization of arbitrary workflows represented with the modeling framework. The resulting Pareto front of the multi-criteria optimization provides the decision makers a set of optimal workflows from which they can choose to optimally fit the current demands. The advantages of the herein presented approach are highlighted with a real world use case from an ongoing research project.
\end{abstract} 
\keywords{ADAPT, Multi-objective optimization, Decision making, Assembly task optimization}

\section{Background and Motivation}
The latest industrial revolution, called Industry 4.0, bundles a multitude of different trends and technologies towards the ongoing digital transformation of industrial manufacturing. This transformation is due to various aspects such as increasing product variety and short product life cycles, triggered by consumer demands and the economic interest of manufacturers. The manufacturing industry reacts accordingly by either focusing on highly automated assembly lines, or highly customized manual assembly following the economy of scope. This situation is especially challenging for small and medium-sized manufacturers, whose small batch size production and limited funds inhibits an amortization of a higher grade of automation, excluding them from many benefits of the ongoing digital transformation. The FELICE project\footnote{FlExible assembLy manufacturIng with human-robot Collaboration and digital twin modEls, see https://www.felice-project.eu/} aims to provide a new solution for digitized and flexible assembly lines to increase both the level of automatism and the ability to react to a dynamic and changing environment. Premise of this solution is a formal representation of any given workflow containing all necessary assembly tasks executed during manufacturing, for which the ADAPT\footnote{ADAPT: Asset-Decision-Action-Property-RelaTionship} modeling approach~\cite{lindorfer2018adapt} is utilized. This formal representation of a workflow provides the opportunity for computational parsing, monitoring, execution and optimization. This formalization further allows the integration of adaptive elements (such as height-adaptable workstations improving the ergonomics for human workers~\cite{froschauer2021human}) and \emph{collaborative robots}, or in short \emph{cobots} (which are tasked to reduce the physical strain of human workers), in workflows. These additional assets can be utilized at various degrees by manipulating the given, formalized workflows. The inclusion or exclusion of adaptive elements results in a change of different, often rivaling key performance indicators (KPIs), such as the total duration of the workflow execution (makespan), ergonomic penalty, or physical strain of the human worker. As the available adaptive elements and the corresponding KPIs are use case dependent, the need for a generic approach arises. This paper aims to provide such generic framework for the multi-objective optimization of ADAPT workflows based on generic user-defined optimization criteria. 

\section{Methodology and Technologies}
In the scope of the herein presented approach, the problem of optimizing workflows is treated as a multi-objective optimization problem~\cite{deb2014multi}, utilizing ADAPT~\cite{lindorfer2018adapt} for the formal representation of workflows and the HeuristicLab framework~\cite{wagner2014} for the optimization of such workflows. For this purpose we employ implementations of a set of suitable algorithms, such as NSGA-II~\cite{deb2002fast} and NSGA-III~\cite{yuan2014improved}. The resulting Pareto front and the corresponding models can be seen as a prescriptive solution that provides the assembly line operator with different optimal suggestions for fine-tuning of the assembly line according to current demands. For example, a tight deadline might result in the prioritization of a minimal makespan, while normally the minimization of the resulting physical strain of the workflow on the human worker might be the highest priority. Depending on the number of different optimization criteria, the problem might also transmute into the domain of many-objective optimization~\cite{manyobjective2008}, which is generally the case as soon as four or more objectives are present. In this scenario, other algorithms such as the NSGA-III are better suited. As HeuristicLab includes optimization algorithms for both of these categories, the approach is agnostic of the underlying number of optimization criteria.

\subsection{ADAPT modeling approach}
\begin{figure}[!htb]
\includegraphics[width=\textwidth]{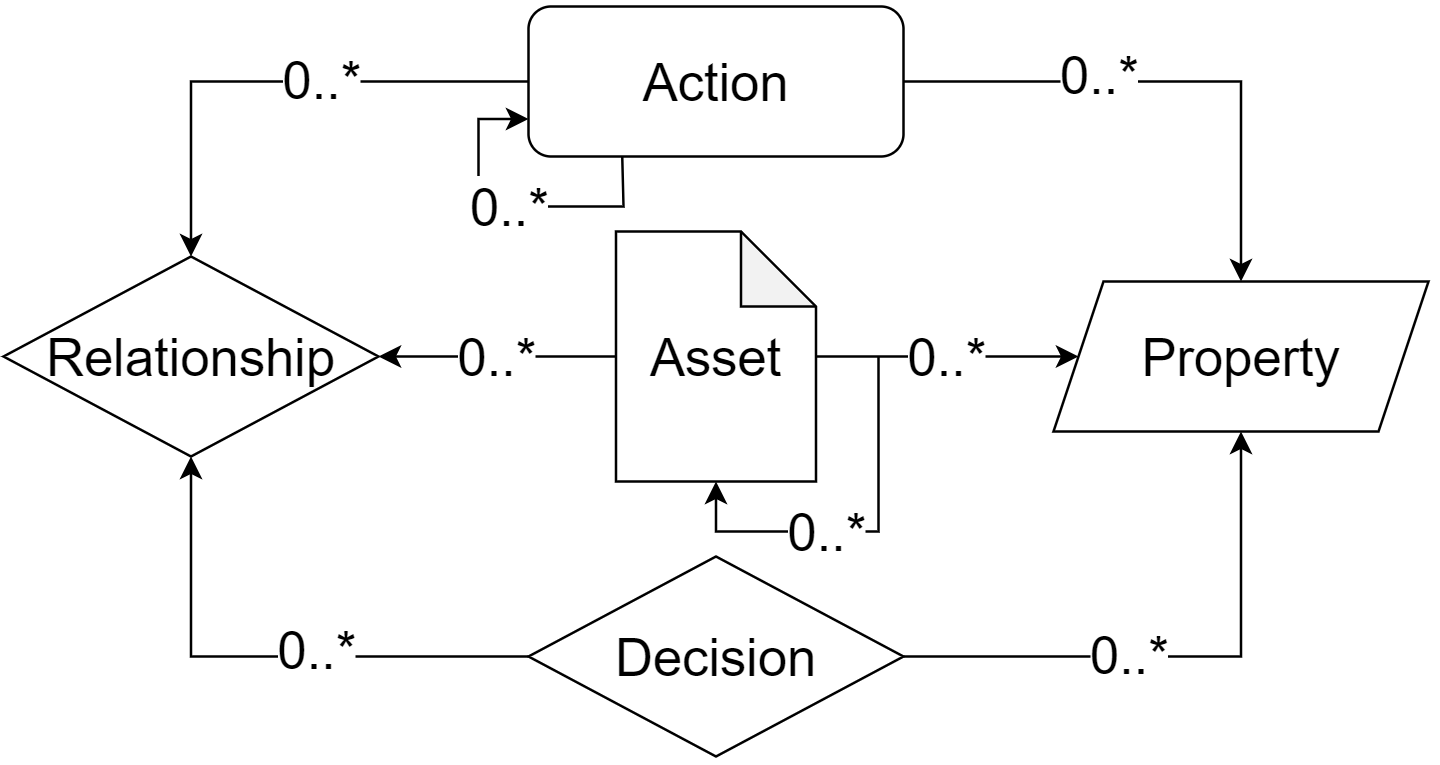}
\centering
\caption{An overview of the ADAPT elements and their interconnection.}
\label{fig:adapt_overview}
\end{figure}

Basis for any optimization is a formal representation of the underlying problem. For workflows, e.g. representing assembly tasks, BPMN~\cite{white2004introduction} is a well established standard and the theoretical origin of other approaches, such as the ADAPT~\cite{lindorfer2018adapt} modeling approach. An overview of the core elements is illustrated in Figure~\ref{fig:adapt_overview} and a more detailed explanation is given as follows.

\begin{itemize}
  \item The \textbf{action}-element is used to model tasks (such as grab, move, screw, ...), which can also be seen as skills~\cite{ferreira2012skill} being required from the current executing entity (human worker, robot, ...). Actions can also be combined into composite actions. 
  \item An \textbf{asset} describes accompanying information which can be both produced (documents) or consumed (positional data, robot instructions, ...) by a workflow.  
  \item A \textbf{decision} element extends workflows with the capability to react according to environmental conditions by creating a fork point for conditional workflows. Although decisions are modeled offline, they can access connected \emph{assets} and react according to present conditions (camera data, positional data, input from user, ...). 
  \item \textbf{Relationship}-elements are used to define existing relationships between the previously defined elements. They usually include successor-relationships between \emph{actions} (order of execution) and include/produce-relationships between \emph{actions} and \emph{assets}.  
  \item \textbf{Properties} allow the definition of additional, generic information for the elements \emph{decision}, \emph{asset} and \emph{action}. 
\end{itemize}

By defining such an ADAPT meta-meta-model, domain-specific workflows can be created and reused for various business processes. Although not strictly limited to a specific field of application, the primary use case for the ADAPT modeling approach is the design of workflows representing assembly tasks for production lines. The ADAPT modeling approach is further accompanied by three tools\footnote{https://sar.fh-ooe.at/index.php/de/downloads/category/3-hcw4i}, called WORM (Workflow Modeler, a graphical WYSIWYG-frontend), HCW4i Runtime (Engine for workflow execution) and HCW4i Visualisation (Frontend for the visualisation of workflow execution and accompanying information, including visual feedback and input for decisions). The underlying workflows themselves are persisted as xml files and therefore human-readable and platform-independent which allows for simple transferal between different frameworks such as WORM and HeuristicLab.

\subsection{HeuristicLab}
HeuristicLab is an open source framework for heuristic optimization\footnote{https://github.com/heal-research/HeuristicLab}. The developers focused on the creation of a paradigm-independent, flexible and extensible design, mainly achieved with a generic plugin-infrastructure. It features a variety of optimization algorithms and problem definitions for different domains. The available optimization algorithms encompass different types such as trajectory and population based, classification and regression, and single and multi-/many-objective. Of particular interest for the proposed solution are three core components of the HeuristicLab plugin-infrastructure, namely the problems, encodings, and  multi-/many-objective algorithms. As indicated by the very name of the plugin-infrastructure, new problems can be defined and included in HeuristicLab, as long as they are derived from the predefined, generic interface called IProblem. Besides the generic IProblem, several more specific, but easier to use base classes are available, such as the MultiObjectiveBasicProblem, which can be implemented to create new multi-objective problems with a specific encoding such as binary vector or permutation encoding. One especially interesting feature with regard to encoding is the availability of a class called MultiEncoding, which acts as a wrapper for a list of encodings, providing the ability to utilize several potentially different encodings for a problem. 

\subsection{Integration}
\begin{figure}
\includegraphics[width=\textwidth]{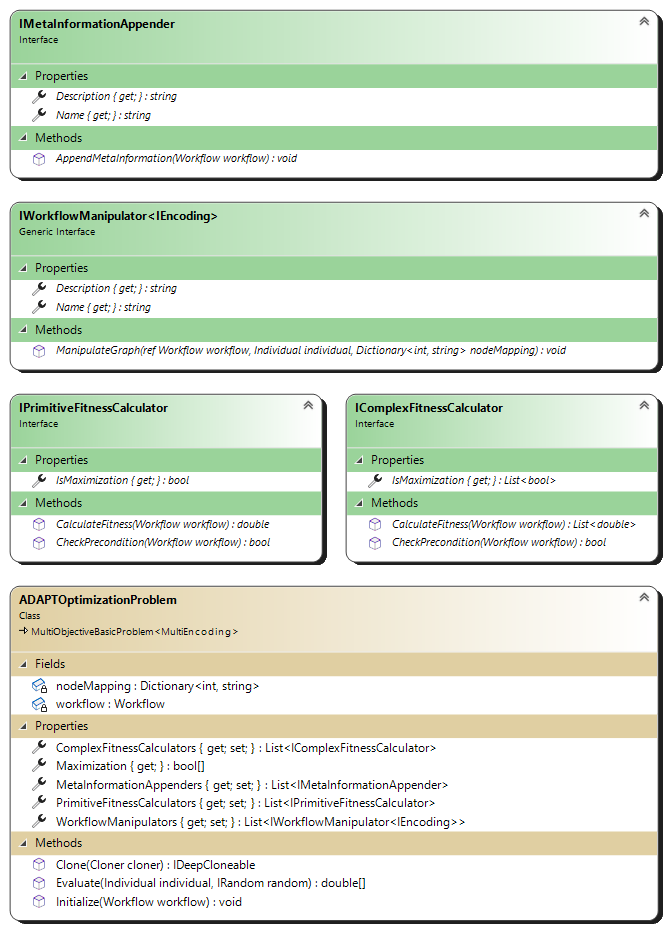}
\centering
\caption{An overview of the new classes and interfaces within the proposed framework.}
\label{fig:felice_interfaces}
\end{figure}

Foundation of the multi-criteria optimization of ADAPT workflows is the creation of a new problem definition in HeuristicLab, named ADAPTOptimizationProblem, which is derived from MultiObjectiveBasicProblem$ < $MultiEncoding$ > $. In addition to this problem definition, a new set of interfaces is introduced, which must be implemented according to a given use case. An overview of these interfaces is depicted in Figure~\ref{fig:felice_interfaces}, a detailed explanation is given as follows. 

\begin{itemize}
  \item IMetaInformationAppender: The purpose of the IMetaInformationAppender is to embed additional (meta-)information into a given workflow. The recommended way is to add new properties to the existing set of actions, assets and decisions. Although there are no restrictions in terms of quantity of properties, each implementation of IMetaInformationAppender should preferably add only one specific property, which is indicated by the Name and explained in the Description.
  \item IWorkflowManipulator$ < $IEncoding$ > $: The IWorkflowManipulator manipulates a workflow according to a given instance of an IEncoding. The manipulation can range from simple property manipulation (change value of properties according to RealVectorEncoding) up to a complete reordering of the whole workflow (permutation of the execution order of actions by changing relationships according to a permutation encoding). Similar to the IMetaInformationAppender, the IWorkflowManipulator should also only represent a single concern which is again indicated in the Name and Description. As an implementation of this interface potentially alters the elements of the workflow, a mapping between the initial enumeration of the actions in the original workflow and their ID is provided. 
  \item IPrimitiveFitnessCalculator: Represents a fitness/objective of a workflow, which is calculated with the CalculateFitness method. The IsMaximization defines whether the objective should be minimized or maximized. The method CheckPrecondition can be used to validate if a workflow is eligible for fitness calculation (implementations of IWorkflowManipulator might create infeasible workflows).
  \item IComplexFitnessCalculator: Each IPrimitiveFitnessCalculator traverses the workflow and calculates a fitness, but sometimes a single workflow traversal is sufficient to calculate a number of fitness values. To improve runtime performance, the IComplexFitnessCalculator is defined, allowing to return a number of fitness values at once.
  \item ADAPTOptimizationProblem: The ADAPTOptimizationProblem bundles all relevant information for the optimization of workflows, including a list of the previously defined interfaces. As the problem adheres to the required interfaces, it can be easily integrated in HeuristicLab and solved with any of the available and compatible algorithms.
\end{itemize}

The general procedure of ADAPTOptimizationProblem starts by initializing the four previously defined lists of interfaces. Afterwards, the Initialize method is called, which initializes the workflow and enumerates all action nodes, creating a mapping between the initial order and the ID. The corresponding encodings of the IWorkflowManipulator implementations are used to initialize the MultiEncoding of the problem. After this initialization, the workflow is manipulated by calling the AppendMetaInformation method from each IMetaInformationAppender. As soon as these steps are executed, the optimization can start and guides the search for optimal solutions. Solution candidates are generated by executing the ManipulateGraph methods of the IWorkflowManipulator implementations on the workflow. Finally, the fitness is calculated by executing the IPrimitiveFitnessCalculator and IComplexFitnessCalculator implementations on the manipulated workflow.
  
\section{Use case}
A major goal of the aforementioned FELICE project is the utilization of collaborative robots to reduce the amount of physical strain on the human worker. One of the basic assumptions is that each of the defined actions can be executed by either the human worker or the cobot, each with different implications. As the cobot has to ensure the safety of the human workers, the maximum velocity of movements is limited, hence most actions will take longer when executed by the cobot. In contrast to the presumably slower cobot, the human worker experiences fatigue, especially during execution of unergonomic actions. Both of these aspects, duration and ergonomic impact, can be quantified by various methods. In the scope of the FELICE project, we focus on the Methods-time measurement (MTM) system~\cite{maynard1948methods} for the estimation of the duration of each action in seconds and the MURI Analysis~\cite{papoutsakis2022detection} for the ergonomic penalty expressed as ordinal values (one, two or three, the higher the better). In accordance with the proposed interfaces and this simple but illustrative use case, we can now define four new IMetaInformationAppender classes, appending properties called ExecutionTimeHuman, ErgonomicPenaltyHuman, CobotExecutionTime and IsCobotUtilized, representing the aforementioned metrics including a flag on whether a cobot is utilized or not. A new IWorkflowManipulator$ < $BinaryVectorEncoding$ > $ (length of the encoding equals to the number of actions in the workflow) class is designed to alter the IsCobotUtilized flag. An implementation of the IComplexFitnessCalculator interface traverses the workflow and aggregates the corresponding execution times from either the human worker or cobot and the ergonomic penalty or zero (depending on whether the IsCobotUtilized flag is set to false for the corresponding action). This results in a set of pareto-optimal solutions in terms of makespan and ergonomic penalty. We're currently in the process of integrating the proposed framework in a real world assembly line setup, allowing us to gather data and to validate the framework in the near future. 

\section{Discussion and Conclusion}
This paper presents a new, generic approach for the multi-criteria optimization of workflows representing assembly tasks. The novelty of the approach lies within the combination of two existing frameworks for workflow creation and optimization. As shown in the use case, this approach can be used to model multi-criteria optimization problems in manufacturing and is utilized in the FELICE project. One current limitation is the absence of precedence rules within the IMetaInformationAppender and IWorkflowManipulator interfaces. This might lead to unintended behaviour if the same property is manipulated more than once in one iteration. The currently fixed-length encoding and enumeration according to the original elements hinders insertion and removal of elements and requires workarounds in the code. These issues are currently under investigation and the solution will be improved in the scope of the FELICE project.  

\subsection*{Acknowledgments}
\vspace{-2mm}
\begin{figure}
  \centering
  \begin{minipage}[c]{0.22\textwidth}
    \centering
    \includegraphics[width=0.8\textwidth]{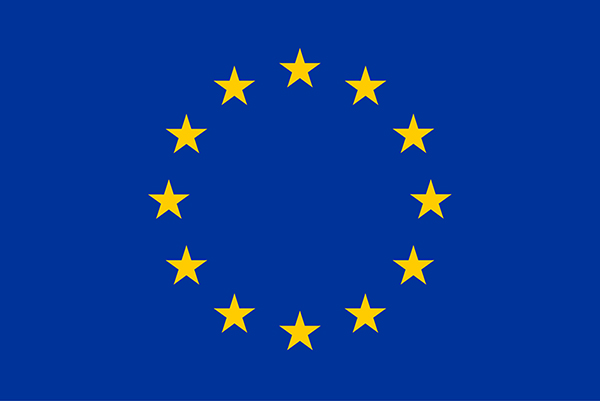}
    \end{minipage}
  \begin{minipage}[c]{0.7\textwidth}
    This project has received funding from the European Union’s Horizon 2020 research and innovation programme under grant agreement No 101017151.	
  \end{minipage}
\end{figure}

\bibliographystyle{splncs04}
\bibliography{Holzinger}

\end{document}